# The role of radiation-induced segregation in defect-phase formation in Ni-Ge and Ni-Si alloys

*Amit Verma[1,2,\*], Yen-Ting Chang[1], Marie Charpagne[1], Pascal Bellon[1], Robert S. Averback[1]*

[1] Department of Materials Science and Engineering, The Grainger College of Engineering, University of Illinois Urbana-Champaign, 1304 W. Green St., Urbana, IL, 61801, USA

[2] BARC, Mumbai, India

## Abstract

The interactions between chemical phase fields and structural defects play a key role in the properties of alloys. We illustrate the importance of these interactions in driven alloys, where defects are continuously being created, with particular focus on systems where radiation-induced segregation occurs. Specifically, we compare the microstructural evolution in undersaturated Ni-Si and Ni-Ge alloys during both 100 keV He and 2 MeV Ti irradiations. While the equilibrium phase diagrams of these systems are similar, and both systems show strong radiation-induced segregation, the evolving defect structures are remarkably different. Ni-Si reveals a high density of Frank loops, while Ni-Ge shows a complex array of dislocations. Moreover, a $Ni_3Ge$ precipitate shell is observed to coat He bubbles, while no segregation of Si is observed at such bubbles. We explain these differences in behaviors to solute drag by interstitial fluxes in Ni-Si vs solute drag by vacancy fluxes in Ni-Ge.

Radiation induced segregation (RIS) has long been a concern in the design of structural nuclear materials owing to its impact on both phase stability and defect structures in alloys [1-3]. Advection of solute to dislocations and grain boundaries, for example, can strongly influence their mobilities and structure, and thereby alter mechanical properties [4, 5], while the depletion of solute in the grain interior can reverse the radiation resistance imparted by such solute, as illustrated by void swelling arising from the loss of Si from stainless steels [6-8]. While past work had largely focused on the kinetic details of RIS in attempts to identify which solutes will segregate and the underlying mechanisms involved [9-12], more recent attention has turned to understanding how solute controls the evolution of defect structures [13-17] and how specific defect structures impact phase formation [18, 19]. A convenient description for the latter process has been developed for near equilibrium systems by in the form of defect-phases, where the chemical phase field is strongly coupled to the underlying defect structure [20-23]; however, in systems undergoing RIS, developing defect structure can be strongly influenced by addition/loss of solute, while the distribution of solute and precipitate phases will respond to the evolving defect structure. In the present work, we illustrate this added complexity by considering the defect structure and radiation-induced segregation in two quite similar undersaturated alloys, Ni-Si and Ni-Ge.

Ni-Si and Ni-Ge are selected for this comparative study primarily since their phase diagrams are quite similar, with the L1$_2$ (Cu$_3$Au) structure, or γ', being the first of several intermetallic phases to form with increasing solute concentration. Both alloys, moreover, reveal significant solute enrichment at defect sinks due to RIS [24], although the mechanism of RIS in these two alloys appears to be different. For Ni-Si, the mechanism of RIS involves the formation of highly mobile interstitial-solute (i-Si) complexes [25], {100} mixed dumbbells, which have a binding energy of 0.79 eV [12]. The vacancy-solute (v-Si) binding energy, in contrast, is only ~ 0.095 eV [26]. For Ni-Ge, the i-Ge mixed dumbbell interstitial also appears stable and highly mobile [27, 28], but its binding energy is much lower than i-Si, 0.22 eV [26] and quite similar to that of the v-Ge complex, 0.19 eV [25]. The present work illustrates how these different point-defect-solute interactions in dilute, nominally single-phase alloys lead to very different driven defect-phase structures.

The samples for the present studies include both bulk alloys and nanocrystalline thin films. The bulk samples were prepared by standard methods, as described in the Supplementary Material, and they had compositions of Ni-7.5 at.% Ge and Ni-8.0 at.% Si. The nanocrystalline samples were grown using magnetron co-sputtering of pure targets to form Ni–9.8at.%Ge and Ni-8.5 at.%Si on sapphire substrates held at 450 °C during growth. Both alloys grew with a columnar grain structure with in-plane grain sizes of ~ 700 nm and ~ 260 nm, respectively. The total oxygen content in these films was < 0.05. at.% as determined by Atom Probe Tomography (APT) analysis. Additional details are reported in the Supplementary Material.

We begin discussion of RIS in Ni-Ge and Ni-Si with the comparison of the driven defect-phase structures in the two alloys using 100 keV He$^+$ irradiation at 450 °C, as illustrated for the bulk specimens of Ni-Ge, and Ni-Si in Figs. 1 and 2, respectively. The STEM-BF micrographs, taken on the <001> zone axis, illustrate the dramatically different defect structures created in the two alloys: the Ni-Ge sample mostly contains a complex array of dislocations within the irradiated zone and a small number of unfaulted dislocation loops, as indicated by yellow arrows in Fig. 1. In stark contrast, the Ni-Si sample reveals predominantly, faulted dislocation loops. Both samples also contain a high density of He bubbles, ~ $10^{17}$-cm$^{-3}$, with mean diameters of ~ 2 nm. As shown in the Supplementary Material, Table S1, the bubble size appears to be somewhat larger in the Ni-

Si samples, a factor of ~ 2. Also shown in Figs. 1 and 2 are STEM-EDS heat maps showing the local concentrations of solute. Both figures clearly illustrate solute segregation at dislocations. For the Ni-Si system, this is revealed most clearly by the EDS line scan denoted by arrow #2, which crosses two loops whose habit plane lies normal to the viewing direction. For Ni-Ge the complex network of dislocation lines makes this identification more difficult, but comparison of Figs. 1(b) and 1(c), again shows this behavior. The absolute solute concentrations at defects shown in the heat maps are under-represented in these figures owing to averaging through the thickness of the TEM specimens, ~ 50 nm. The electron diffraction patterns in Figs. 1(a) and 2(a), however, clearly reveal that the $\gamma'$ phase is formed in both alloys; this is also suggested by the scattered "red" voxels in the concentration heat maps representing solute concentrations of ~25 at.%. The defect microstructures observed in the Ni-Ge and Ni-Si thin-film (nanocrystalline) samples are quite similar to those in their respective bulk alloys; although the dislocation loops form in the Ni-Si nanocrystalline samples appear relatively smaller in size. This behavior is attributed to the higher density of grain boundaries in the nanocrystalline sample, which act as competing sinks for defects. Segregation at GB's in the two nanocrystalline alloys also showed similar behavior, as illustrated in the Supplementary Material, Figs. S3 and S4. Notably, segregation is strongest at the triple junctions in both alloys.

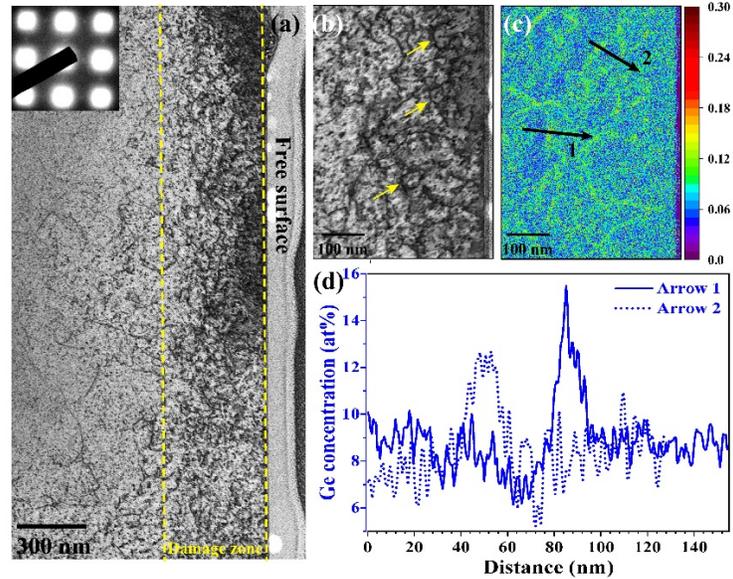

**Fig.1.** Microstructure of bulk Ni-7.5at%Ge alloy irradiated with 100 keV He$^+$ ions at 450°C for 1.95 dpa dose. (a) On <001> zone axis low magnification STEM-BF micrograph (b)-(d) represent high magnification STEM-BF, STEM-EDS heat map of Ge and EDS line scan along two arrows passing through dislocations shown in (b) and (c), respectively. A <001> zone axis SAED pattern is shown in the inset of (a). Yellow arrows in (b) point to unfaulted dislocation loops.

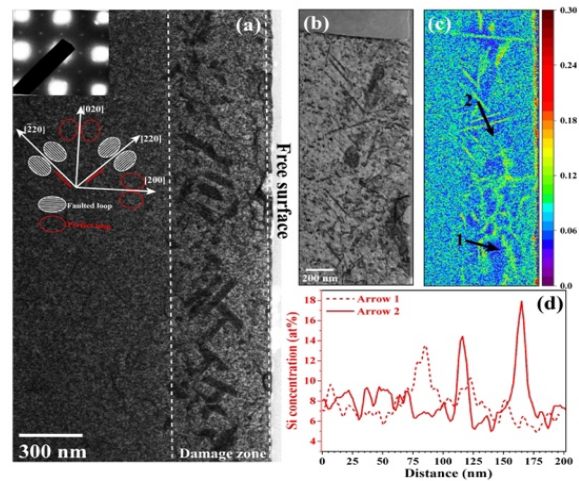

**Fig.2.** Microstructure of bulk Ni-8at%Si alloy irradiated with 100 keV He$^+$ ions at 450°C for 1.95 dpa dose. (a) On <001> zone axis low magnification STEM-BF micrograph depicting faulted dislocation loops in the irradiated zone of Ni-8at%Si alloy as schematically illustrated in the inset of the figure. A <001> zone axis SAED pattern is shown in the inset of (a). (b)-(d) High magnification STEM-BF, STEM-EDS heat map of Si and EDS lines scan along two arrows passing through edge-on dislocation loops marked in (c), respectively. STEM-BF micrograph in (b) was obtained in the <110> zone axis of the FCC crystal.

The He bubbles in these samples are also potential sinks for point defects, and therefore, possible sites for segregation. As noted above, the sizes and densities of He bubbles are not dramatically different in these two alloys; the segregation behavior, however, is, as illustrated in Figs. 3 and 4 for the thin film samples. Fig. 3(a-c) employs STEM and 4-D STEM imaging to locate a high-angle GB and two nearby large He bubbles in Ni-Ge. This region was subsequently examined by APT using a proxigram analysis [29], illustrating that the GB is decorated with a high density of He bubbles. The two larger He bubbles, moreover, are enclosed by a highly enriched Ge shell, Fig. 3(d,e,f), while Fig. 3(g) presents an atom-density heat map, revealing the low atomic density associated with the bubble. Lastly, Fig. 3(h) shows that the Ge concentration at locations greater than ~ 2 nm away from the bubble is close to the average Ge alloy concentration in the grain interior after irradiation, 9.5.at% Ge, but that it increases rapidly as the location approaches the He bubble, plateauing at ~ 25 at.%, indicating a $\gamma'$ shell has formed around the bubble. The corresponding analysis for Ni-Si, Fig. 4, shows quite different behavior. While both alloys show significant segregation of solute at the GB's (Figs. S3 and S4), the $\gamma'$ precipitates near the GB show no evidence of a significant presence of He (Fig. 4c-4e). The proxigram analysis, Figs.4(g,h), corresponding to Si rich and He rich regions identified in the APT reconstruction of a second APT specimen (Fig. 4(f)), moreover, demonstrates (i) that near a representative $\gamma'$ precipitate there is no excess He and (ii) the Si concentration decreases, rather than increases, near the edge of the He bubble. Furthermore, TEM-dark field (TEM-DF) micrograph of the needle sample reveals a nearly continuous film of $\gamma'$ phase particles along the twin boundary (Fig. 4c) with a composition of approximately 25 at.% Si, as evidenced by the Si heat map shown in Fig. 4d. Lastly, we note that the integral amount of solute lost from solution due to RIS is quite similar for the two alloys, the Ge concentration in solution is reduced from 9.8 at.% Ge to 9.5at.% during irradiation, whereas the Si concentration is reduced from 8.5 at.% Si to 8.2 at.% Si, see Table S2 in the Supplementary Material.

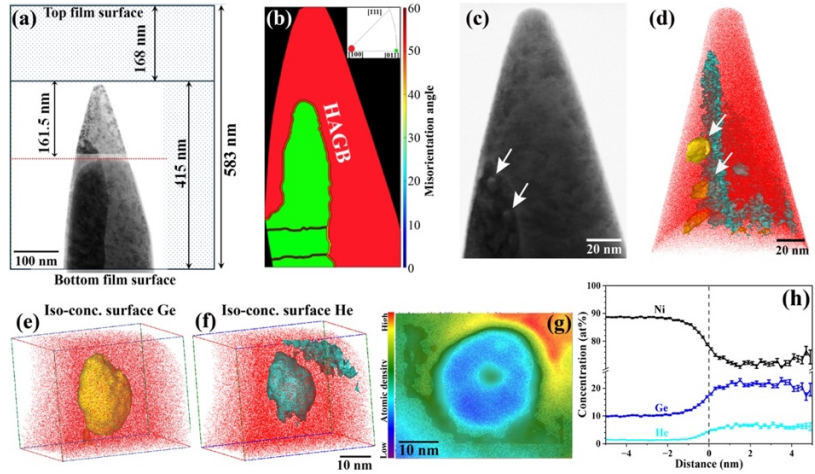

**Fig. 3**. APT analysis of the Ni$_3$Ge phase particles formed in nanocrystalline Ni-8.5Ge sample after irradiation with 100 keV He ions at 450°C for 1.95 dpa dose. (a) shows overlaid STEM-BF micrographs of the APT needle sample before and after APT experiment, and the red-dotted line demarcates the analysed region of the needle. (b) 4D-STEM IPF map depicting presence of high angle grain boundary near the apex of the APT needle. (c) HAADF-BF micrograph revealing presence of two He bubbles at the grain boundaries in a region marked by two white arrows. (d) APT reconstructed volume of the needle comprising nickel atoms (red) and iso-concentration surfaces of Ge (yellow) and He (green), depicting location of the grain boundary by green color iso-surface and locations of the He bubble by white arrows. (e-f) demarcate a Ni$_3$Ge phase particle within a small reconstructed volume delineated by iso-concentration surfaces drawn with Ge and He. (g) density heat map of Ni+Ge in a thin square slab passing through the middle of the precipitate shown in (e) and (f) revealing relatively a lower atomic density within the core than in the shell. (h) proxigram of Ni$_3$Ge phase particle shown in (e).

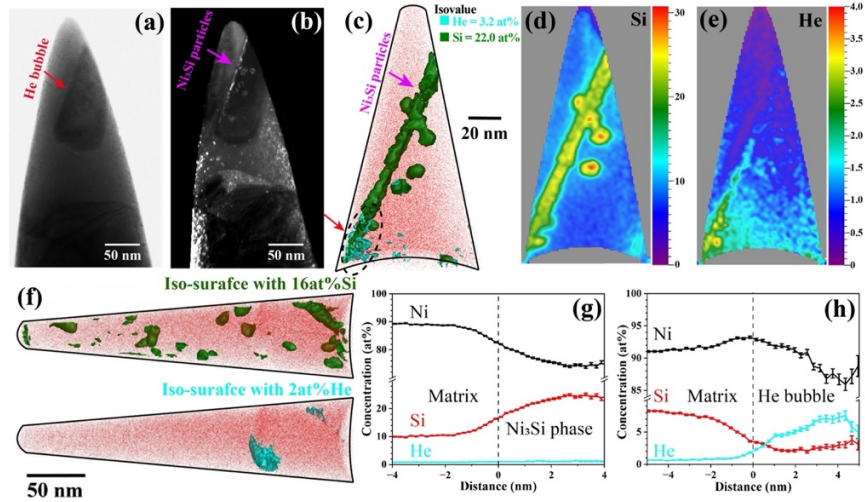

**Fig. 4.** (a)-(c) TEM-BF, TEM-DF and APT reconstruction of Ni-8.5at%Si sample after irradiation with 100 keV He ions at 450°C for 1.95 dpa dose. (a) presence of He bubbles at the twin boundary is marked by red arrow (b) TEM-DF of nano-sized particles of the Ni$_3$Si phase particles at the grain boundary marked by pink arrow; (c) APT reconstruction delineating presence of continuous film of Ni$_3$Si phase particles at the twin boundary by blue iso-concentration surface of Si; (d-e) show heat map of Si and He within a thin slab containing a section of twin boundary and Ni$_3$Si particles, respectively; (f) APT reconstruction volume of a second sample comprising nickel atoms (red) and iso-concentration surfaces of Si (green color) and He (cyan) delineating intragranular Ni$_3$Si phase particles and one He bubble; (g-h) show proxigrams from one Ni$_3$Si phase particle and one He bubble, respectively.

The last part of these experiments reports on the defect-phase behavior in the bulk alloys irradiated with 2 MeV Ti$^+$. Most notably, comparison of Fig. 5 with Figs. 1 and 2 reveals that despite the very different primary recoil spectra between 2 MeV Ti and 100 keV He, and the lack of He bubbles, the defect states produced by the two irradiations are quite similar, *viz.* Ni-8at.%Si is again comprised mainly of faulted dislocation loops, while Ni-7.5at.%Ge acquires a complex network of dislocations. The segregation behavior, moreover, is also qualitatively quite similar although segregation at defects appears somewhat weaker despite the higher dose (in dpa). Ni-8at.% Si shows significant segregation at the free surface and γ′ particles at dislocation loops (See Supplementary Material Fig. S5). The segregation at the GB's appears weak, (black arrow #2 in Fig. 5. While this GB is located near the end of the damage distribution, analysis of the second

GB (pink arrow) shows nearly the same excess of Si and neither GB revealed γ'. The Ni–7.5 at.% Ge sample similarly exhibited weaker radiation-induced segregation (RIS) at the grain boundaries for the Ti irradiation, with no evidence of γ' phase formation anywhere in the microstructure. While the irradiation dose with Ti for this sample was roughly half that used for Ni-Si, it was ~ 4 times larger than the He dose. Thus, while the defect structures are very similar in both alloys for He and Ti irradiations, RIS appears weaker for the Ti irradiation in both alloys. This latter finding, however, is completely consistent with past results on the effects of recoil spectrum on RIS efficiency [30].

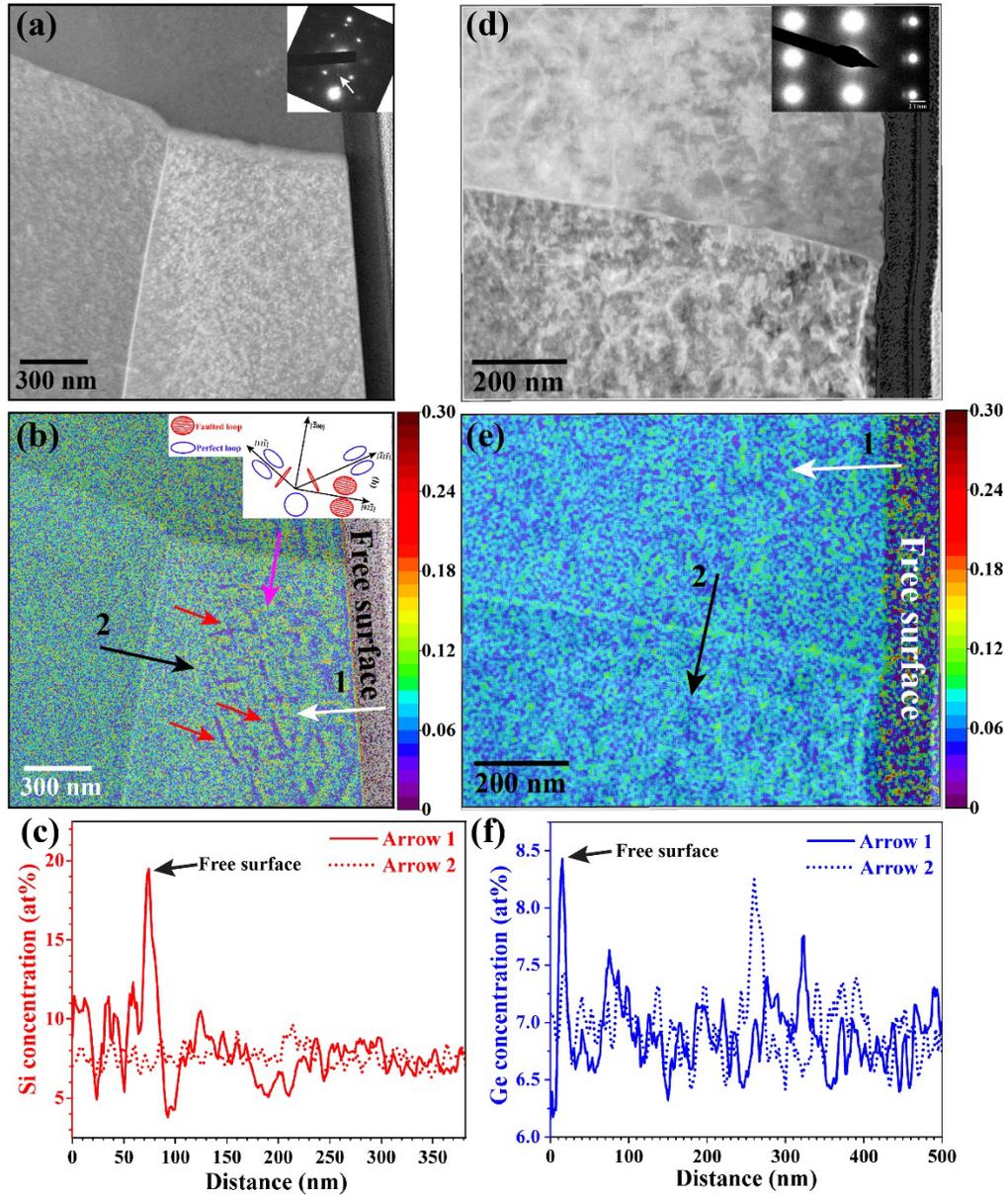

**Fig.5.** (a) and (d) On zone axis STEM-HAADF micrographs of Ni-8at%Si (<011>; 10 dpa) and Ni-7.5at%Ge (<112>; 4.6 dpa) alloys samples irradiated with 2 MeV Ti$^+$ ions at 450°C, respectively; (b-e) STEM-EDS heat map overlaid with STEM-BF micrograph of Ni-8at%Si and Ni-7.5at%Ge alloys, respectively; (c) and (f) EDS line scan along two arrows passing through free surfaces (arrow 1) and across grain boundaries (arrow 2). EDS line scan along arrow 1 reveals stronger RIS at the free surface in the Ni-8Si sample and weaker RIS in the Ni-7.5at%Ge sample (free surface position marked by black arrow in (c) and (d)). EDS line scan across grain boundaries exhibits weaker RIS in both alloys. SAED pattern of each alloy is shown in the inset of (a) and (b). Red color arrows in (b) mark location of dislocation loops with habit plane perpendicular to projection plane. Presence of superlattice reflection of the Ni$_3$Si phase in <110> zone axis SAED pattern confirms precipitation of the Ni$_3$Si phase in the Ni-Si alloy due to RIS, while no superlattice reflections of the Ni$_3$Ge phase in the <112> zone axis SAED pattern in the Ni-Ge alloy.

The present results for the influence of solute on the defect structures, which is in qualitative agreement with past work using neutron irradiation at 300 °C on 2 at.% alloys to 0.4 dpa [17], can largely be understood from past diffuse x-ray scattering measurements. These studies show that interstitial clusters in electron- or neutron-irradiated Ni and Ni-1 at.% Ge are highly mobile and form loops that grow quickly during annealing, while in Ni-1at.%Si the Ni-Si mixed dumbbells are highly mobile, but that small clusters of them are not. Consequently, interstitial atoms in pure Ni and Ni-Ge can migrate rapidly to the surface and GB sinks, allowing vacancy loops to form and grow. Moreover, these intrinsic loops can subsequently unfault to form dislocation arrays [7, 8, 31, 32], which are decorated with Ge solute and γ' precipitates. In Ni-Si, the interstitial clusters are immobilized, i.e., they become the slower moving defect, and thus they increase their concentration relative to vacancies by enabling cluster growth by additions of newly created interstitial-Si dumbbells, This scenario is supported by atomic resolution STEM-BF and STEM-HAADF images and diffraction patterns in the Supplementary Materials (Fig. S5) and ref. [33] that illustrate that γ′ has formed at the loop edge but not in the loop interior; it also explains why the EDS heat maps in Fig. 2(c) show regions depleted of Si near the Frank loops.

The failure to observe segregation, or γ′, at He bubbles in Ni-8.5 at.% Si appears to arise from the high He pressure within the bubble. For example, if it is assumed that all of the He is contained in bubbles, then using a bubble density of $1 \times 10^{17}$-cm$^{-3}$ and an average diameter 5 nm, (see Table S1) the bubble pressure at 773 K is ~ 10 GPa. This introduces a pΔV penalty of roughly 0.6 eV for adding an interstitial atom to the bubble. For Ni-Ge the binding energy of vacancies to solute is comparable to that of interstitials, consequently the bubbles provide an excellent sink for v-Ge complexes. In this same light, the smaller size and higher density of He bubbles in Ni-Ge could arise from the Ni$_3$Ge γ′ shell surrounding the He bubbles, which presents an obstacle to He entering the bubble. While this latter possibility introduces a potential means to control the growth of He bubbles, an issue in Ni-based alloys [34], uncertainties in evaluating the He bubble pressure in the current study prevent a more quantitative conclusion regarding this possibility.

## Acknowledgments

Work by A. V., Y.-T. C, M. C., P. B. and R. S. A. was supported by the U.S. Department of Energy, Office of Science, Basic Energy Sciences, under Award No SC0019875. This work made use of the Illinois Campus Cluster, a computing resource that is operated by the Illinois Campus Cluster Program (ICCP) in conjunction with the National Center for Supercomputing Applications (NCSA) and which is supported by funds from the University of Illinois at Urbana-Champaign. Materials' characterization and irradiation were performed at the Materials Research Laboratory (MRL) Central Research Facilities, University of Illinois. Atom probe tomography was performed at the MRL using a CAMECA LEAP 5000-XS instrument purchased with support from the NSF under Grant No. DMR-1828450. This work was performed, in part, at the Center for Integrated Nanotechnologies, an Office of Science User Facility operated for the U.S. Department of Energy (DOE) Office of Science by Los Alamos National Laboratory (Contract 89233218CNA000001) and Sandia National Laboratories (Contract DE-NA-0003525). We also acknowledge helpful technical discussions with Dr. Yongqiang Wang.

Supplementary Material for

# The role of radiation-induced segregation in defect-phase formation in Ni-Ge and Ni-Si alloys

*Amit Verma[1,2,*], Yen-Ting Chang[1], Marie Charpagne[1], Pascal Bellon[1], Robert S. Averback[1]*

[1] Department of Materials Science and Engineering, The Grainger College of Engineering, University of Illinois Urbana-Champaign, 1304 W. Green St., Urbana, IL, 61801, USA

[2] BARC, Mumbai, India

**Materials and Methods**

**Fabrication of nanocrystalline thin films:** A DC magnetron sputtering unit housing sputtering targets of pure nickel (99.999%), pure silicon (99.99%) and pure germanium (99.99%) (from Angstrom Engineering) was used to grow nanocrystalline thin film samples of Ni-Si alloys and Ni-Ge alloys. The film growth was carried out in a sputter deposition chamber with a substrate platform stage modified to accommodate miniature SiN heating plates from Bach Resistor Ceramics Gmbh, Germany. Thin-film samples were deposited on a sapphire substrate kept in direct contact with the heating plate and maintained at ~ 450°C. The base pressure of the deposition chamber was ~ $4 \times 10^{-8}$ torr; the deposition was carried out while dynamically pumping ultrahigh purity argon at a pressure of ~$1.8 \times 10^{-3}$ torr. The films were grown by co-depositing pure Ni and Ge targets to form Ni–Ge alloys, and pure Ni and Si targets to form Ni–Si alloys, to produce films of ~ 578 nm and ~ 407 nm thicknesses, respectively. Rutherford backscattering spectrometry (RBS) and transmission electron microscopy (TEM) analysis of samples prepared in the cross-sectional plane were used to determine the thicknesses of two layers as well as the nominal Si and Ge concentration in the as-grown samples of Ni-Si and Ni-Ge alloys. The films grew with a columnar grain structure with compositions of 9.8 at.% Ge and 8.5 at.% Si; the grain sizes were nm ~ 700 nm ~ 260, respectively.

**Fabrication of bulk alloys:** Bulk alloys were prepared to provide a reference for the irradiation response of Ni-Si alloys with large grain sizes. Melting and casting were employed to prepare a 40 g button of Ni-8at%Si alloy from elemental charges of nickel with 99.99% purity and silicon with 99.999% purity using non-consumable arc melting furnace. The melting was carried out after first evacuating the furnace chamber to ~$10^{-6}$ torr and flushing it with high-purity argon gas for three repetitions, and then backfilling with Ar to ~ 0.5 atm. The alloy button was homogenized by melting it six times, flipping it after each melting. Next, the alloy button was sealed in a quartz ampule after evacuating to a pressure of ~ $10^{-6}$ torr and backfilling with high-purity helium gas to a pressure of ~160 mm of Hg, where it was homogenized at 1250°C for 24 hours. Lastly, the alloy was cold rolled to a thickness of ~ 1 mm and subjected to a recrystallization treatment at 750°C for 2 h under vacuum at a pressure < $10^{-7}$ torr. TEM-BF micrographs illustrate the pre-irradiation microstructures in Fig. S1.

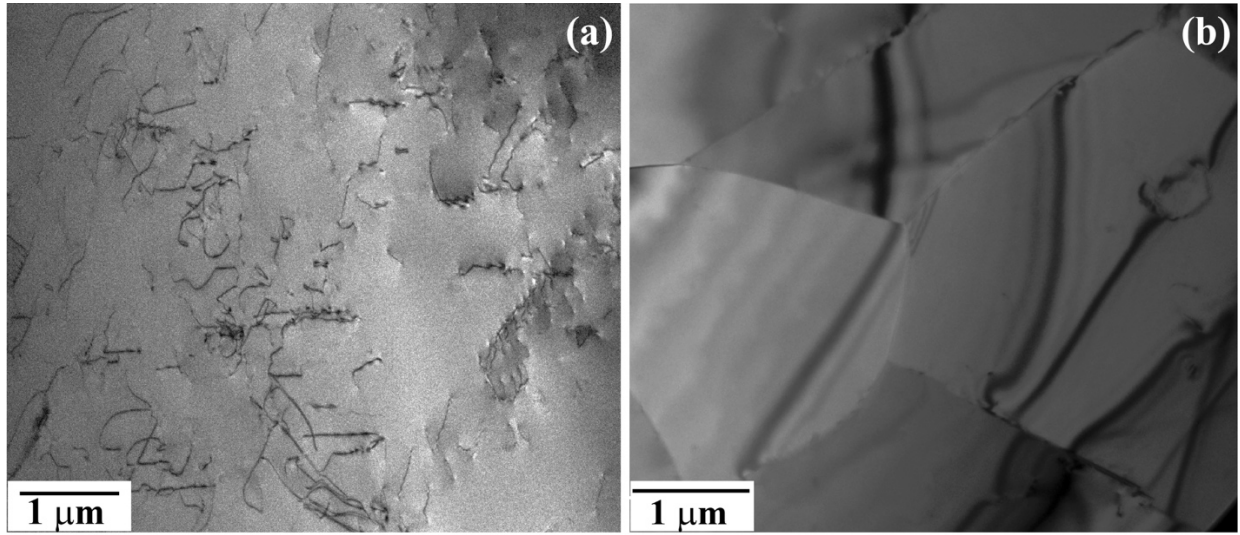

Fig. S1.(a) TEM-BF micrographs of initial microstructure of arc-melted and recrystallized samples of (a) Ni-7.5at%Ge alloy illustrating extensive presence of dislocations; (b) Ni-8at%Si alloy revealing comparatively cleaner microstructure. Dislocation density of ~ $2.9 \times 10^9 /cm^2$ were found in the Ni-7.5at%Ge alloy.

**Irradiation:** Light ion irradiations of both bulk and thin film samples were carried out at the LANL using 100 keV $He^+$ ions employing 1cm circular beam that raster at 1 kHz frequency across 85 mm x 85mm area. Irradiations were carried at temperatures of 450 ± 5 °C to a total dose of 1.98 dpa in a high vacuum chamber with a background pressure ~ $4 \times 10^{-7}$ torr. Heavy ion irradiations of bulk samples were carried out at the UIUC-MRL Accelerator Laboratory using 2 MeV $Ti^+$ ions. Heavy ion irradiations were carried at temperatures of 450 and 530 ± 5 °C in a high vacuum chamber with a background pressure < $8 \times 10^{-8}$ torr. The ion flux during the irradiation was ~ $1.7 \times 10^{12}$ ions $cm^{-2}$ $s^{-1}$, yielding a damage rate in the range from 1.6 - $2.3 \times 10^{-3}$ dpa $s^{-1}$, depending on the depth below the surface. Dose rates were obtained using the NRT model as implemented in SRIM. Total damage levels were 4.6 and 10 dpa. The addition of implanted titanium was negligible for samples investigated at a depth ≈ 300 nm. Damage and implantation profiles appear in Fig. S2

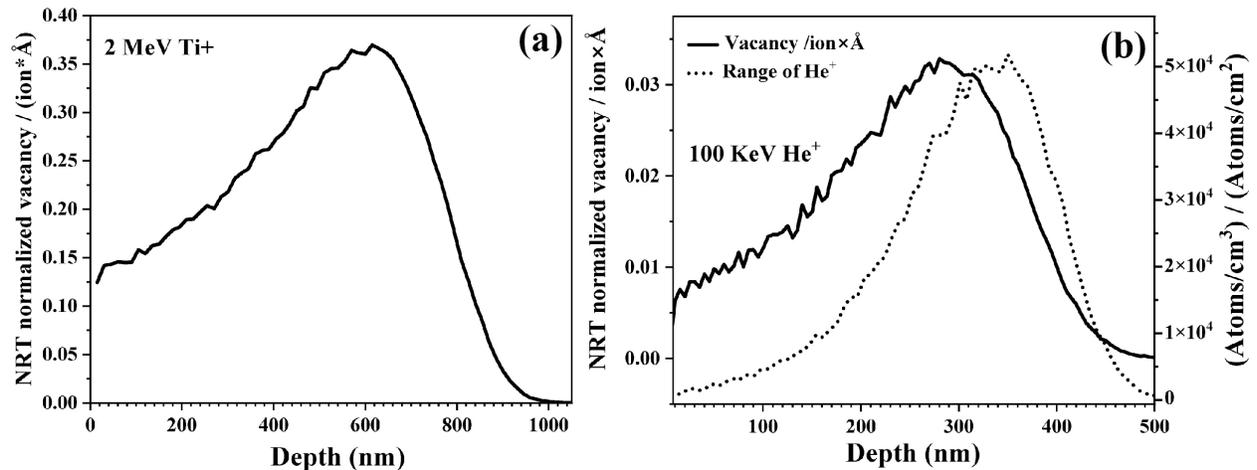

**Fig. S2** SRIM damage profiles for Ni-8%X (X=Si/Ge) alloys: (a) 2MeV Ti$^+$ ions; and (b) 100 keV He$^+$ ions. Fig. (b) also includes implantation profile of He$^+$ ions in the sample.

**Characterization:** Transmission electron microscopy (TEM), scanning TEM (STEM) and atom probe tomography (APT) were used to characterize the sample microstructures. TEM lamellas were lifted out from cross-sectional planes of the irradiated samples of Ni-Si and Ni-Ge bulk alloys and their thin film samples. In addition, plan-view TEM lamellas were lifted out at a depth of ~250 nm from thin film samples using FIB-lift out techniques using a Thermoscientific Scios 2 FIB-SEM. APT specimens were fabricated on 5-post stainless steel grids using well-established APT specimen preparation techniques in a FIB-SEM [1]. APT needles were prepared in a cross-section view for Ni-9.8at%Ge thin film sample and in plan view of Ni-8.5at%Si thin film samples by lifting out lamellas at a depth of ~250 nm . A Jeol 2010 TEM equipped with LaB$_6$ filament and operating at 200 KeV was used for microstructural characterization using bright field (BF) and dark field (DF) imaging modes in conjunction with selected area electron diffraction (SAED). Thermoscientific TALOS F200X STEM operating at 200 keV and Thermoscientific THEMIS Z STEM operating at 300 keV, both equipped with BF, High Angle Annular Dark Field (HAADF) and Energy Dispersive Spectroscopy (EDS) detectors were also used for microstructural characterization of irradiated samples. APT analysis was performed using a CAMECA straight path Local Electrode Atom Probe LEAP-5000XS. Quantitative analysis of the local chemical composition of samples was performed using energy dispersive spectroscopy (STEM-EDS) and APT. APT experiments were performed both using laser pulsing and voltage pulse mode. APT data was acquired in the laser pulsing mode employing laser energies of 40-70 pJ; pulse frequency of 333 kHz; and detection rate of 0.5%. Sample temperature during each measurement was maintained at 30 K. The mass spectra of each measurement produced golden hits of more than 92% and background noise smaller than 25 ppm/nsec. Prior to APT measurements, needle samples were analysed in STEM using BF and HAADF imaging modes to determine the shape parameters of the specimen (apex radius and half shank angle). Needle samples were also characterized by 4D-STEM to obtain grain orientation map and to determine the misorientation angles of grains near the tip of the needle samples. The 4D-STEM data was collected in the TALOS F200X STEM using an EMPAD (Electron Microscopy Pixilated Array Detector), a 128 × 128 direct electron detector with a high dynamic range of 1,000,000:160. A convergence angle of 1 mrad was used, which corresponds to a probe size of ≈7 nm and the region of interest was divided into a 128 × 128 pixel array. A total of 16384 data sets of real space and reciprocal space information were collected, and subsequently processed using an open source 4D-STEM code, py4DSTEM, which is a Python package. To evaluate the grain boundary misorientations, an open-source MATLAB toolbox (MTEX), was used for quantitative texture analysis. Data reconstruction and analysis were carried out using Image Visualization and Analysis (IVAS) software module embedded in the APT Suite 6.1. The 3D reconstruction was optimized by combining specimen shape parameters and grain geometry information obtained from the STEM investigation of needle samples prior to APT measurement.

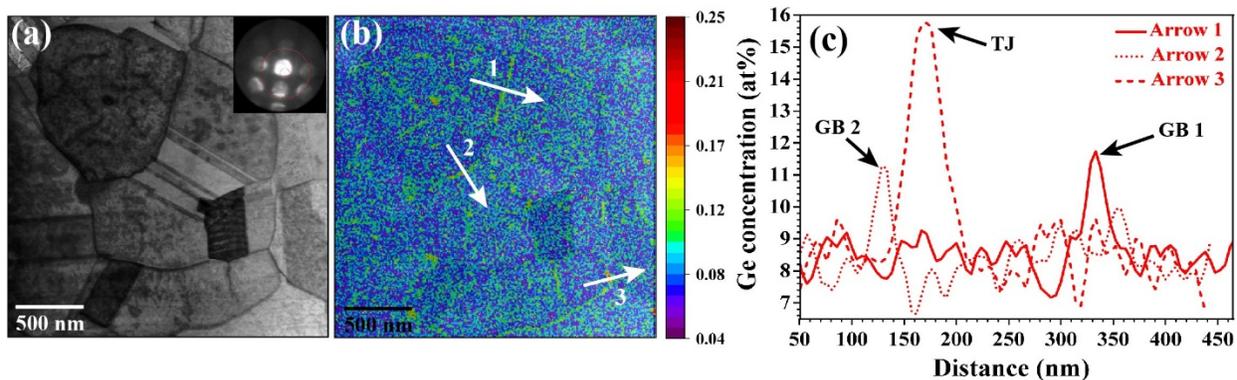

**Fig. S3.** (a) and (b) show an on-zone <001> axis STEM-BF micrograph and a STEM-EDS heat map of Ge of a nanocrystalline Ni-9.8 Ge alloy, prepared in a plan view by FIB lift-out of a TEM lamella at a depth of ~ 250 nm from the free surface. SAED pattern is shown as inset of (a) and 3 arrows drawn on (b) mark locations of the grain boundaries (GB, arrow 1 and 2) and tri-junction (TJ, arrow 3) for solute concentration analysis along the arrows. Fig. (c) shows concentration profiles of Ge along three arrows marked in (b).

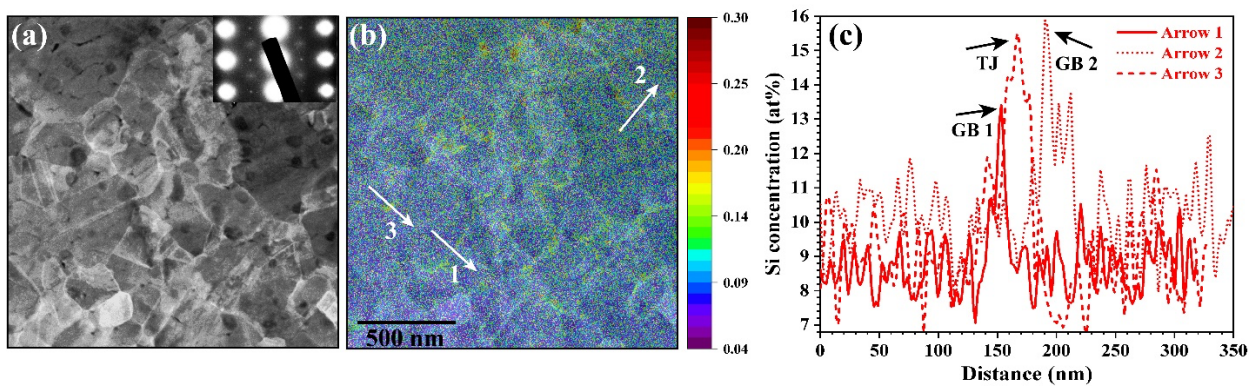

**Fig. S4.** (a) and (b) show a STEM-BF micrograph and a STEM-EDS heat map of Si of a nanocrystalline Ni-8.5 Si alloy, respectively, prepared in a plan view by FIB lifting out TEM lamella at a depth of ~ 250 nm from the free surface. Inset of (a) shows <112> zone axis SAED pattern confirming formation of γ′ phase particles by presence of superlattice reflection at (110) positions. Fig. (b) mark locations of the grain boundaries (GB, arrow 1 and 2) and tri-junction (TJ, arrow 3) for solute concentration analysis along the arrows. Fig. (c) shows concentration profiles of Si along three arrows marked in (b).

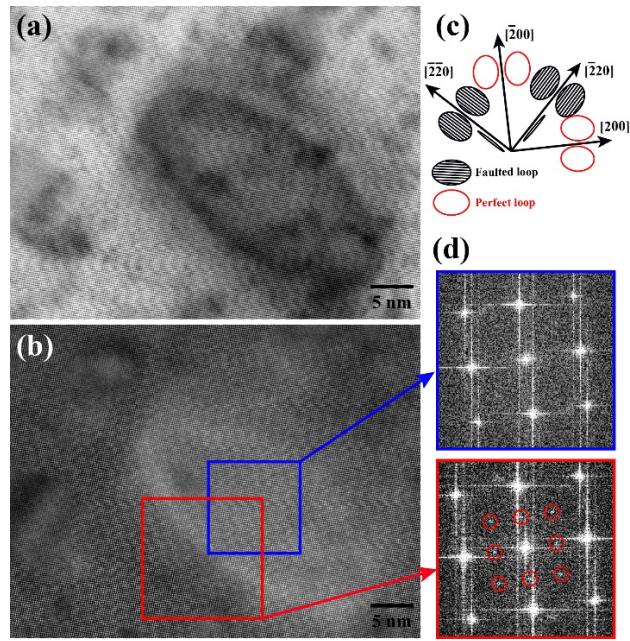

**Fig. S5** (a) Atomic resolution STEM-BF image showing unfaceted Frank loop in Ni-8at.%Si bulk alloy after 2 MeV Ti irradiation to 4.6 dpa at 530 °C; (b) STEM-HAADF image showing Si segregation in the form of γ′ precipitates at loop edge as evident by presence of superlattice reflections in the electron diffraction obtained via fast Fourier transformation of a region enclosed by a red box (containing loop edge), while no superlattice reflection of a region enclosed by a blue box (loop core). The shape of the faulted loop observed in the STEM-BF image is consistent with the predicted morphology of the Frank loop projected along the <001> direction of the FCC crystal as depicted in (c).

**Table S1: Size and density of He bubbles in Ni-Si and Ni-Ge after 100 keV He irradiation to 1.9 dpa**

|  | Average size of He bubble | Density of He bubble /cm$^3$ |
|---|---|---|
| Ni-8.5at%Si-Thin film | 5.0 ± 3.4 nm | 3.5 × 10$^{16}$ bubbles/cm$^3$ @ 170 nm depth from free surface |
|  | 5.4 ± 3.7 nm | 5.6 × 10$^{16}$ bubbles/cm$^3$ @ 220 nm depth from free surface |
|  | 6.4 ± 4.5 nm | 5.9 × 10$^{16}$ bubbles/cm$^3$ @ 300 nm depth from free surface |
| | 5.4 ± 1.3 nm | He bubbles at the grain boundary |
| Ni-8at%Si-bulk alloy | 2.4 ± 1.7 nm | $1.5 \times 10^{17}$ bubbles/cm$^3$ @ 170 nm depth from free surface |
| | 2.5 ± 1.7 nm | $2.5 \times 10^{17}$ bubbles/cm$^3$ @ 250 nm depth from free surface |
| | 2.4 ± 1.7 nm | $3.1 \times 10^{17}$ bubbles/cm$^3$ @ 300 nm depth from free surface |
| Ni-9.8at%Ge-Thin film | 2.1 ± 0.2 nm | $2.5 \times 10^{17}$ bubbles/cm$^3$ @ 170 nm depth from free surface |
| | 2.1 ± 0.2 nm | $2.4 \times 10^{17}$ bubbles/cm$^3$ @ 250 nm depth from free surface |
| | 2.2 ± 0.3 nm | $3.0 \times 10^{17}$ bubbles/cm$^3$ @ 300 nm depth from free surface |
| | 7.4 ± 2.4 nm | He bubbles at the grain boundary |
| Ni-7.5at%Ge-bulk alloy | 1.7 ± 0.3 nm | $1.2 \times 10^{17}$ bubbles/cm$^3$ @ 170 nm depth from free surface |
| | 1.7 ± 0.3 nm | $2.2 \times 10^{17}$ bubbles/cm$^3$ @ 250 nm depth from free surface |
| | 1.7 ± 0.3 nm | $2.3 \times 10^{17}$ bubbles/cm$^3$ @ 250 nm depth from free surface |

**Table S2: Ge and Si in solution in thin film samples before and after irradiation**

|  | As prepared thin film samples | After irradiation at 450°C (1.9dpa) |
|---|---|---|
| **Ni-8.5at%Si** | Si= 8.5at% | Si = 8.2at% |
| **Ni-9.8at%Ge** | Ge = 9.8at% | Ge = 9.5 at% |